\documentclass[%
 reprint,
 superscriptaddress,
 twocolumn, 
 amsmath,amssymb,
 aps,
]{revtex4-2}
\usepackage{todonotes}
\usepackage{breakurl}
\usepackage[colorlinks = true,
            linkcolor = blue,
            urlcolor  = blue,
            citecolor = blue,
            anchorcolor = blue]{hyperref}
\usepackage[normalem]{ulem}
\usepackage{graphicx}
\usepackage{dcolumn}
\usepackage{bm}
\usepackage{braket}
\usepackage{glossaries}
\usepackage{subfigure}
\newacronym{CDW}{CDW}{charge-density-wave}
\newacronym{QMC}{QMC}{quantum monte carlo}
\newacronym{DFPT}{DFPT}{density functional perturbation theory}
\newacronym{DFT}{DFT}{density functional theory}

\usepackage{ulem}

\newcommand{\expect}[1]{{\langle #1 \rangle}} 

\makeatletter
\newsavebox{\@brx}
\newcommand{\llangle}[1][]{\savebox{\@brx}{\(\m@th{#1\langle}\)}%
  \mathopen{\copy\@brx\kern-0.5\wd\@brx\usebox{\@brx}}}
\newcommand{\rrangle}[1][]{\savebox{\@brx}{\(\m@th{#1\rangle}\)}%
  \mathclose{\copy\@brx\kern-0.5\wd\@brx\usebox{\@brx}}}
\makeatother

\newcommand{\COMMENTED}[1]{}

\begin{document}

\preprint{}

\title{Metal-insulator transition and quantum magnetism in the SU(3) Fermi-Hubbard Model:
Disentangling Nesting and the Mott Transition}
\author{Chunhan Feng}
\affiliation{Center for Computational Quantum Physics, Flatiron Institute, 162 5th Avenue, New York, New York, USA}
\author{Eduardo Ibarra-Garc\'ia-Padilla}
\affiliation{Department of Physics and Astronomy, San Jos\'e State University, San Jos\'e, California 95192, USA}
\affiliation{Department of Physics and Astronomy, University of California, Davis, California 95616, USA}
\author{Kaden R. A. Hazzard}
\affiliation{Department of Physics and Astronomy, Rice University, Houston, TX 77005, USA}
\affiliation{Rice Center for Quantum Materials, Rice University, Houston, TX 77005, USA}
\affiliation{Department of Physics and Astronomy, University of California, Davis, California 95616, USA}
\author{Richard Scalettar}
\affiliation{Department of Physics and Astronomy, University of California, Davis, California 95616, USA}
\author{Shiwei Zhang}
\affiliation{Center for Computational Quantum Physics, Flatiron Institute, 162 5th Avenue, New York, New York, USA}
\author{Ettore Vitali}
\affiliation{Department of Physics, California State University Fresno, Fresno, California 93720, USA}

\date{\today}

\begin{abstract}

We use state-of-the-art numerical techniques to compute ground state correlations
in the two-dimensional SU(3) Fermi Hubbard model at $1/3$-filling, modeling fermions with three possible spin flavors moving on a square lattice with an average of one particle per site. 
We find clear evidence of a quantum critical point separating a non-magnetic uniform metallic phase
from a regime where long-range `spin' order is present. In particular, 
there are multiple successive transitions to 
 states with regular, long-range alternation of the different flavors, whose symmetry changes as the interaction strength increases. In addition to the  
 rich quantum magnetism,  
 this important physical system  
 allows one to study integer filling and the associated Mott transition disentangled from nesting, in contrast to the usual SU(2) model.
 Our results also provide a significant step towards the interpretation
of present and future experiments on fermionic alkaline-earth atoms,
and other realizations of SU($N$) physics.
\end{abstract}

\maketitle

\noindent
\textit{Introduction}.~The Fermi-Hubbard model (FHM)\cite{hubbard1963electron,gutzwiller1965correlation} is a paradigmatic description of strongly correlated 
materials\cite{rasetti1991hubbard,scalapino1995distinguishing}.
In its original single-band, SU($2$) symmetric form, it exhibits a wealth of physics
including a metal-to-insulator transition at half-filling\cite{gebhard1997metal}, as well as ferro- and antiferro-magnetic orders
across its interaction-filling phase diagram\cite{hirsch1985two,white1989numerical,fazekas1999lecture,SU2-U-n-PhaseDiag-PhysRevResearch.4.013239}.
On a square lattice, upon doping away from half-filling, it also manifests more subtle physics including
strange metallicity, a pseudogap, spin-charge `stripe' domains, and $d$-wave pairing, phenomena central
to the cuprate superconductors\cite{Tasaki1998,Arovas2022,Qin2022rev}.

However, in this geometry and at half-filling
the properties of the SU(2) Hubbard Hamiltonian are determined simultaneously
by the special features of the non-interacting dispersion --- perfect nesting at $k=(\pi,\pi)$ ---
and by the on-site repulsion $U$. 
 This leads to the anomalous feature that the critical interaction strength for the metal-insulator
 transition is $U_c=0$, and a blurring of `Slater' insulating behavior associated with the 
 opening of an antiferromagnetic gap at $(\pi,\pi)$, and the Mott insulator which does not
rely on long range magnetic symmetry breaking but only on the high energy cost for local double occupancy.
 While one can separate these effects, for example
by adding a next-near-neighbor  hopping $t'$, it is hard to tune such parameters in experimental
realizations.

The SU($N$) Hubbard model, which features larger spins and an enhanced symmetry, has been intensively studied \cite{honerkamp2004,Titvinidze2011,Sotnikov2014,Sotnikov2015,HafezTorbati2018,HafezTorbati2019,HafezTorbati2020,Nie2017,gorelik2009,PerezRomero2021,IbarraGarciaPadilla2021}, in part because quantum fluctuations are expected to give rise to a more complex set of low temperature spin structures. A rich phenomenology has been predicted in the ground state in the Heisenberg limit ($U\gg t$ 
 at $\expect{n}=1$)\cite{hermele2009,toth2010,hermele2011,nataf2014,corboz2011,bauer2012,Yamamoto2020,nataf2016,Romen2020,song2013},
 including an antiferromagnetic (AFM)
ground state with a 3-sublattice pattern 
at the wavevector $k = (\pm 2\pi/3, \pm 2\pi/3)$\cite{toth2010,Romen2020}. The physics away from this limit has been less explored, 
with existing results  
largely focused on  fairly high-temperature properties, 
employing 
methods that are uncontrolled at low temperature \cite{Hazzard2012,Bonnes2012,Cai2013,Cichy2016,Golubeva2017,Sotnikov2014,Unukovych2021,Sotnikov2015,honerkamp2004}, or restricted to one dimension \cite{Manmana2011,Xu2018} or half filling for even $N$ \cite{Blumer2013,wang19,Wang2014,Cai2013_2,Assaad2005} (which is challenging to reach due to particle losses in experiments with ultracold atoms in optical lattices).

 Simultaneously, the rapid development of experiments with ultracold alkaline-earth-like atoms (AEAs) that exhibit a natural SU($N$) symmetric interaction\cite{Gorshkov2010,Cazalilla2014,Wu2003,Cazalilla2009,Taie2012,Hofrichter2016,Ozawa2018,Taie2020,Takahashi2022,Tusi2022,Pasqualetti2023}
 provides a concrete and exciting 
 setting to explore these intriguing phases, including the explicit  separation of the  non-generic features of the band structure on a square lattice from the effect of interactions.  
 Ongoing experiments aiming to pair quantum gas microscopes \cite{Gross2017,Bloch2012,Yamamoto2016,Gross2021} with AEAs are expected to soon enable real-space imaging of correlations, which demands a deeper understanding of the order in the SU($N$) Hubbard model.

It is a major challenge to obtain 
reliable answers to the question we address here, namely 
the ground-state properties of the SU(3) FHM on a 2D square lattice at 1/3-filling.
A high accuracy treatment is required in a strongly correlated system,  as well as
large supercell sizes to extract the thermodynamic limit. 
Unlike its SU(2) counterpart, a fermion sign problem \cite{loh1990sign} 
is present in the SU(3) model at $\langle n\rangle=1$.
In this paper we use a novel implementation of the constrained path (CP) auxiliary-field quantum Monte Carlo
(AFQMC) methodology \cite{zhang97} which iteratively refines the constraining Slater determinant \cite{qin16}.
This allows us to quantify 
the location of the metal-to-insulator transition
and determine a phase diagram.
Above a critical interaction strength $U_c$ 
which
is roughly half the bandwidth, the system
develops AFM order. 
We also discover  two intermediate AFM orders (which we label 
3-2 and 3-4 AFM) along the route to the large $U$ Heisenberg limit.
 
\vskip0.10in
\noindent
\textit{Hamiltonian and Methodology.} 
The SU(3) FHM is defined by the Hamiltonian
\begin{equation}\label{eq:Hubbard_N1}
H = -t \sum_{\langle i,j \rangle, \sigma} \left( c_{i \sigma}^\dagger c_{j \sigma}^{\phantom{\dagger}} + \mathrm{h.c.} \right) + \frac{U}{2} \sum_{i,\sigma \neq \tau} n_{i \sigma} n_{i \tau}
\end{equation} 
where $c_{i \sigma}^\dagger$ ($c_{i \sigma}^{\phantom{\dagger}} $) is the creation (annihilation) operator for a fermion with spin flavor $\sigma = A,B,C$ on site $i$ 
on a 2D square lattice. 
$N_s=L_x\times L_y$ denotes the number of lattice sites, $n_{i \sigma}^{\phantom{\dagger}} = c_{i \sigma}^\dagger c_{i \sigma}^{\phantom{\dagger}}$ is the number operator for flavor $\sigma$ on site $i$, 
${\langle i,j\rangle}$
denotes 
nearest-neighbor pairs,
$t$ is the 
hopping amplitude which will serve as the energy unit, and $U$ is the interaction strength.
We study the ground state of the Hamiltonian in Eq.~\eqref{eq:Hubbard_N1} as a function of 
$U$ in the
spin-balanced case at $1/3$-filling 
(where there is an average of one fermion per lattice site),
that is, in the sector of Hilbert space 
with $N_A = N_B = N_c = N_s/3$.

To characterize the ground-state properties, we measure energies, correlation functions, and structure factors.
For example the 
spin-resolved two-body correlation functions,
$\langle n_{i}^{\sigma}  n_{j}^{\sigma'} \rangle \equiv
\langle \, \phi_{0} \, | \, n_{i}^{\sigma}  n_{j}^{\sigma'} \, | \phi_{0} \, \rangle $. 
This is complemented by the corresponding momentum-space measures, for example the 
equal time,
equal species, density structure factor
$S(k)=\frac{1}{3}\sum_{\sigma}S_{\sigma\sigma}(k)=\frac{1}{3 N_s}\sum_{i,j,\sigma} \langle \left(n_i^{\sigma} - \langle n_i^{\sigma} \rangle  \right) \left(n_j^{\sigma} - \langle n_j^{\sigma} \rangle  \right) \rangle 
e^{ik\cdot (r_i-r_j)}$, where $r_i$ and $r_j$ are the coordinates of sites $i$ and $j$.

\begin{figure}[t]
 \includegraphics[width=1\columnwidth]{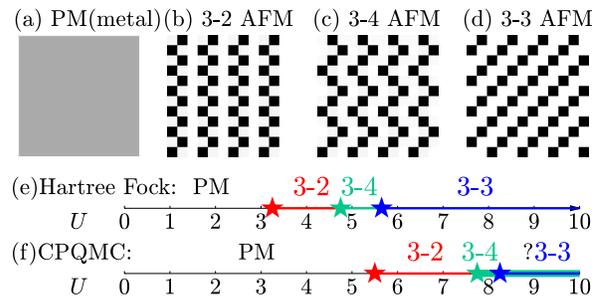}
\caption{Phase diagram of the $1/3$-filling SU(3) Fermi-Hubbard Model. (a-d): 
schematic 
density correlations for one of the three spin species in different phases 
on a $N=12 \times 12$ square lattice. 
(a) At small 
$U<U_{c1}$, 
the system resides in a paramagnetic (PM) phase, as indicated by the uniform gray color map; (b) For $U_{c1}< U < U_{c2}$ a non-uniform pattern 
emerges which has period 3 in one direction and 2 in the other, 
which we 
denote by ``3-2 AFM" phase; (c) $U_{c2} < U < U_{c3}  $ is similarly denoted as ``3-4 AFM"; and (d) finally, for $U>U_{c3}$, the order is a ``3-3 AFM". 
(e) Hartree-Fock phase diagram: 
The quantum phase transition points are
$U_{c1}^{\rm MF} \approx 3.25$, $\, U_{c2}^{\rm MF} \approx 4.75$ and $U_{c3}^{\rm MF} \approx 5.65$.
(f)  AFQMC phase diagram: 
$U_{c1} \approx 5.5$, $\, U_{c2} \approx 7.7$ and 
$U_{c3} \approx 8.3$.
The question mark indicates that, although 3-3 has a slightly lower energy, it 
cannot be clearly distinguished from 3-4 in our calculations.
}
 \label{fig:phase_diagram}
 \end{figure}

We perform unrestricted Hartree-Fock (UHF) calculations 
as preliminary explorations to suggest candidate phases of matter, and to generate trial wavefunctions for AFQMC (see below).
In the UHF treatment we approximate the Hamiltonian as 
 $H_{\rm HF}=\sum_{\sigma}H^{\sigma}_{\rm HF}$,
 where $H^{\sigma}_{\rm HF}=-t \sum_{\langle i,j \rangle} \left( c_{i \sigma}^\dagger c_{j \sigma}^{\phantom{\dagger}}+h.c.\right)
+ U_{\rm eff} 
\sum_{i,\tau \neq \sigma} \left(  \langle  n_{i\tau}\rangle n_{i\sigma}-\frac{1}{2} \langle n_{i\tau}\rangle \langle n_{i\sigma} \rangle \right)$, 
and determine 
the mean-fields $\langle n_{i\sigma}\rangle$ self-consistently from values initialized randomly or 
constructed from 
several possible ordered patterns proposed for the $1/3$ filling SU(3) FHM \cite{Sotnikov2014,Sotnikov2015}. 
Our method reproduces the SU(2) UHF phase diagram 
\cite{hirsch1985two, Xu2011}.
In $H^{\sigma}_{\rm HF}$ above we have allowed an effective interaction 
strength $U_{\rm eff}$, which does not have to equal the ``physical'' $U$. This turns out 
to be important in allowing a self-consistent procedure between UHF 
and AFQMC to determine a best $U_{\rm eff}$ value for producing 
the trial wave function for the CP constraint
\cite{qin16}.
As discussed in more detail later, the four phases depicted in Fig.~\ref{fig:phase_diagram} emerge as 
mean-field ground states for different values of the interaction strength; however, estimates of the critical values of $U$ or even the 
sequence of phases could differ significantly from the exact result.

In order to compute the ground-state properties 
beyond the mean field level, we use the state-of-the-art CP-AFQMC method \cite{zhang97,AFQMC-method-sym-PhysRevB.88.125132}.
AFQMC is a projection quantum Monte Carlo approach, based on the fact that the ground state can be obtained by acting an imaginary time evolution operator on a trial wave function, $\left | \phi_0 \right>  \propto \lim_{\tau \rightarrow \infty} e^{-\tau H }  \left | \phi_T \right>$, as long as the trial wave function is non-orthogonal to the ground state $\braket{\phi_T | \phi_0 } \neq 0$. Within AFQMC, a combination of the Trotter decomposition and the Hubbard-Stratonovich transformation maps the imaginary time evolution onto a random walk in the manifold of Slater determinants, and ground state properties are averages over this random walk.
Observables are measured by back-propagation \cite{zhang97,BP-Purwanto04} in the open-ended random 
walk approach of CP-AFQMC.

\begin{figure}[h]
 \includegraphics[width=1\columnwidth]{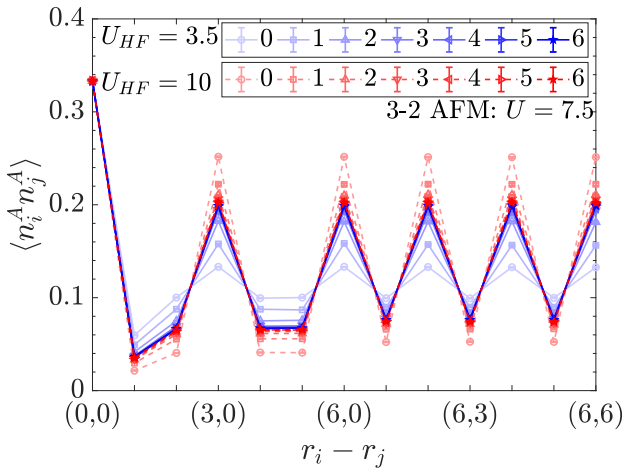}
\caption{Illustration of convergence of real space  density-density correlation.
We have chosen $U_{c1} < U=7.5 < U_{c2}$, where the system is in the 3-2 AF phase.
Spin correlations converge from lower and higher values when the starting point is in the 3-2 phase using
an initial Slater determinant with $U_{\rm eff}=3.5$ and $10$ respectively.  The legend indicates the number of self-consistent iterations.}
 \label{fig:self_consistency}
 
 \end{figure}

AFQMC 
yields exact ground state correlations
in special situations, like the SU(2$N$) Hubbard model at half-filling \cite{Cai2013,Xu2018,wang19}, and electron phonon Hamiltonians including the Holstein \cite{hohenadler19,feng20_2,feng20,chen19} and the Su-Schrieffer-Heeger models \cite{feng22,cai21,xing21,gotz21}.
In the SU(3) model, the infamous fermion sign problem \cite{loh1990sign,troyer2005computational,umrigar2007alleviation,mondaini2022quantum} necessitates a constrained path approximation 
which imposes a sign or gauge condition on the sampling in 
auxiliary-field space \cite{zhang97,chang08} relying on 
a trial wave function. In this work we leverage the recent methodological advances \cite{qin16}, to implement self-consistency loops which minimize the biases related to the trial wave function.
The very promising success of such techniques in the SU(2) model, in both the repulsive \cite{LeBlanc15, zheng17,qin20,qin16_2} and 
attractive regimes \cite{PhysRevLett.128.203201}, together with several cross-checks we did in this work (Supplemental Material, Fig.~\ref{fig:self_consistency}, and 
analysis below), demonstrates the high accuracy of the method.

\vskip0.20in \noindent
\textit{Results.} 
Our main 
 results are summarized in  Fig.~\ref{fig:phase_diagram}, as a
ground-state 
phase diagram of the 1/3-filling ($\braket{n}=1$) SU(3) FHM.  
Panels (a)-(d) illustrate the patterns in the magnetic correlations 
in each of the phases. 
Black and white represent high and low density correlations for a single flavor, 
respectively. 
The many-body phase diagram is given in panel (f). 
(The UHF result is shown in panel (e) for reference and comparison.)
At small $U$, the system is in a paramagnetic (PM) (metal) phase, while at large $U$, multiple occupancies are suppressed and superexchange causes adjacent sites to favor different spin species, and therefore generates an anti-ferromagnetic (AFM) pattern. 
In contrast to the SU(2) case where the AFM correlation 
determines a unique (up to translations) spatial pattern, many patterns are possible for SU(3). We find multiple ordering vectors occur, representing distinct ground states for different $U$ values.
Fig.~\ref{fig:phase_diagram} (b) has a period of $3$ in one direction and $2$ in the other direction, thus it is denoted as a 3-2 AFM. Similarly, patterns (c) and (d) are denoted 3-4 and 3-3 AFM respectively. 

Within the Hartree-Fock treatment, 
we have $U_{c1}^{\rm MF} \sim 3.25$ below which the system is in the paramagnetic phase. 
When $3.25 \lesssim U \lesssim 4.75$, the 3-2 AFM has a lower energy with respect to the other two states, while the 3-4 AFM is the ground state for the intermediate region $4.75 \lesssim U \lesssim 5.65$.  However, the energy gap to other orders in this intermediate regime is relatively small.   The 3-3 AFM is the mean-field ground state for the large U region $U \gtrsim 5.65$. 
The corresponding phase diagram  from AFQMC is shown in 
 Fig.~\ref{fig:phase_diagram}(f). As expected, the metal-insulator phase transition happens at a larger $U_{c1} \sim 5.5 > U_{c1}^{\rm MF} \sim 3.25 $
  than in UHF, 
  reflecting the fact that the latter 
  overestimates the ordering tendency due to its absence of fluctuations. 
As in UHF, AFQMC also gives the 3-3 AFM as the ground state at large $U$, 
consistent with 
the ground state found in prior work in the Heisenberg limit \cite{toth2010}. 
However, the energy of this state, obtained from AFQMC, is nearly degenerate with the 
3-4 pattern for the larger U values investigated here.

 \begin{figure}[t]
 \includegraphics[width=1\columnwidth]{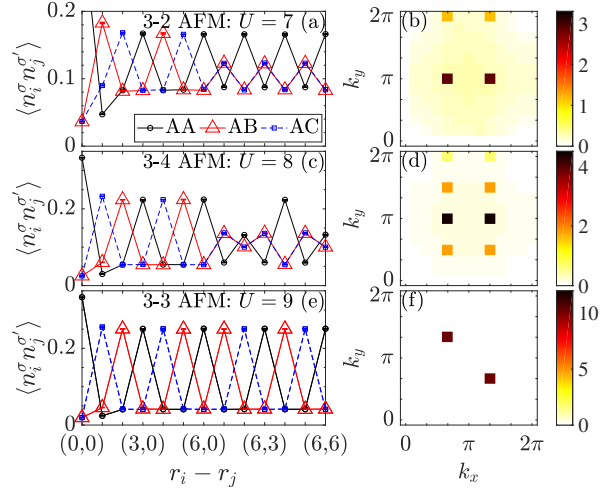}
\caption{(a,c,e): Spin-resolved two-body 
correlations 
$ \langle n_i^{\sigma} n_j^{\sigma^{\prime}} \rangle $ as a function of $r_i-r_j$,
the vector connecting site $i$ and $j$, at $U=7$ (3-2 AFM), $U=8$ (3-4 AFM), $U=9$ (3-3 AFM) given by AFQMC on a $12 \times 12$ lattice. Different colors represent different spin species $\{\sigma,\sigma^{\prime}\}={AA;AB;AC}$. (b,d,f): the single-flavor 
structure factor 
$S_{\sigma\sigma}(k)$
 as a function of momenta $k=(k_x, k_y)$
for the systems in panels a,c, and e, respectively.
}
 \label{fig:dd_corr_real_k_space}
 \end{figure}

  The three AFM phases can be characterized by 
  spin-resolved two-body correlations $\langle n_i^{\sigma}n_j^{\sigma^{\prime}}\rangle$. 
 The correlations are displayed in 
 Fig.~\ref{fig:dd_corr_real_k_space}
 along the real space path  $r_i-r_j=(0,0) \rightarrow (6,0) \rightarrow (6,6)$ in a $12 \times 12$ square lattice. Panels (a,c,e) are for $U=7, 8, 9$,
 and correspond
  to the 3-2, 3-4 and 3-3 AFM phases respectively. Different spin flavor correlations are depicted by different colors. Along the x direction, as seen from separations $(0,0)$ to $(6,0)$, the correlations in all three AFM phases alternate ``ABCABC\ldots" (period 3), while  in the y directions, 3-2, 3-4 and 3-3 AFM have periods 2, 4, 3 respectively. 
The single-flavor structure factor $S_{\sigma\sigma}(k)$, which is a Fourier transform of the 
corresponding real-space two-body correlations, 
is plotted as a function of $k$ 
in Fig.~\ref{fig:dd_corr_real_k_space}(b,d,f). 
The 3-2, 3-4 and 3-3 AFM phases have peaks at distinct ${k}$ points, appropriate to the periodicity of the real space patterns
of Fig.~\ref{fig:phase_diagram}(b-d). 

 \begin{figure}[t] 
 \includegraphics[width=1\columnwidth]{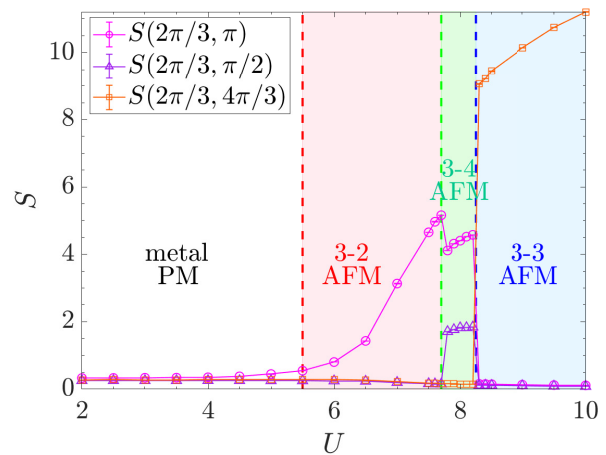}
\caption{The structure factor $S(k)$ 
at characteristic ${k}=(k_x,k_y)$ values of
$(2\pi/3, \pi/2), \, (2\pi/3, \pi), \,  (2\pi/3, 4\pi/3)$ 
 as a function of $U$ 
 on a $12 \times 12$ lattice. The 
AFQMC calculations 
 are started with several different possible trial ground states for 
each $U$.  The one with the lowest energy 
after convergence of the self-consistent iteration 
is shown here.}
 \label{fig:structure_factor}
 \end{figure}

 \begin{figure*}[t]
 \centering
 \subfigure{
 \includegraphics[width=1\columnwidth]{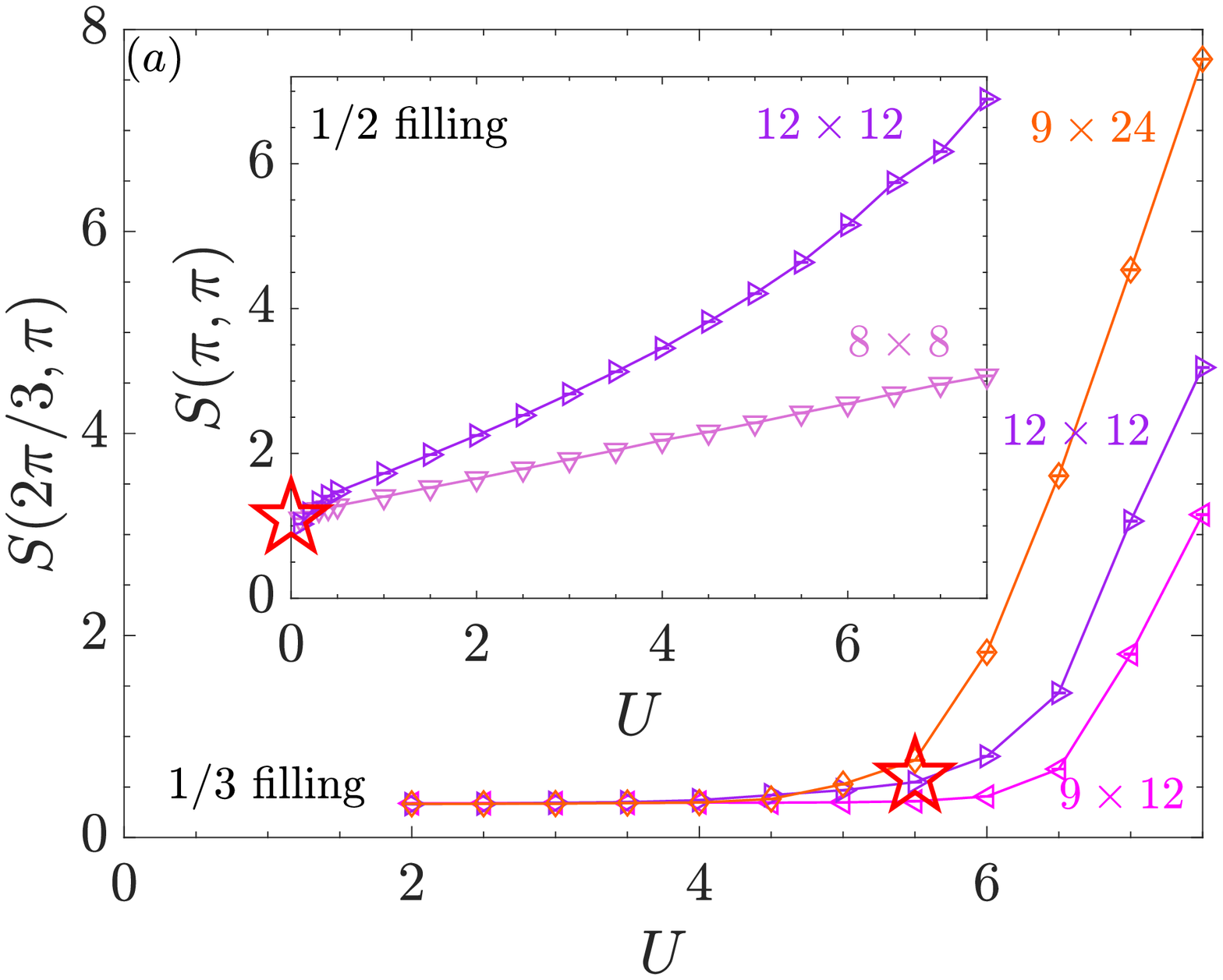}
 \includegraphics[width=1\columnwidth]{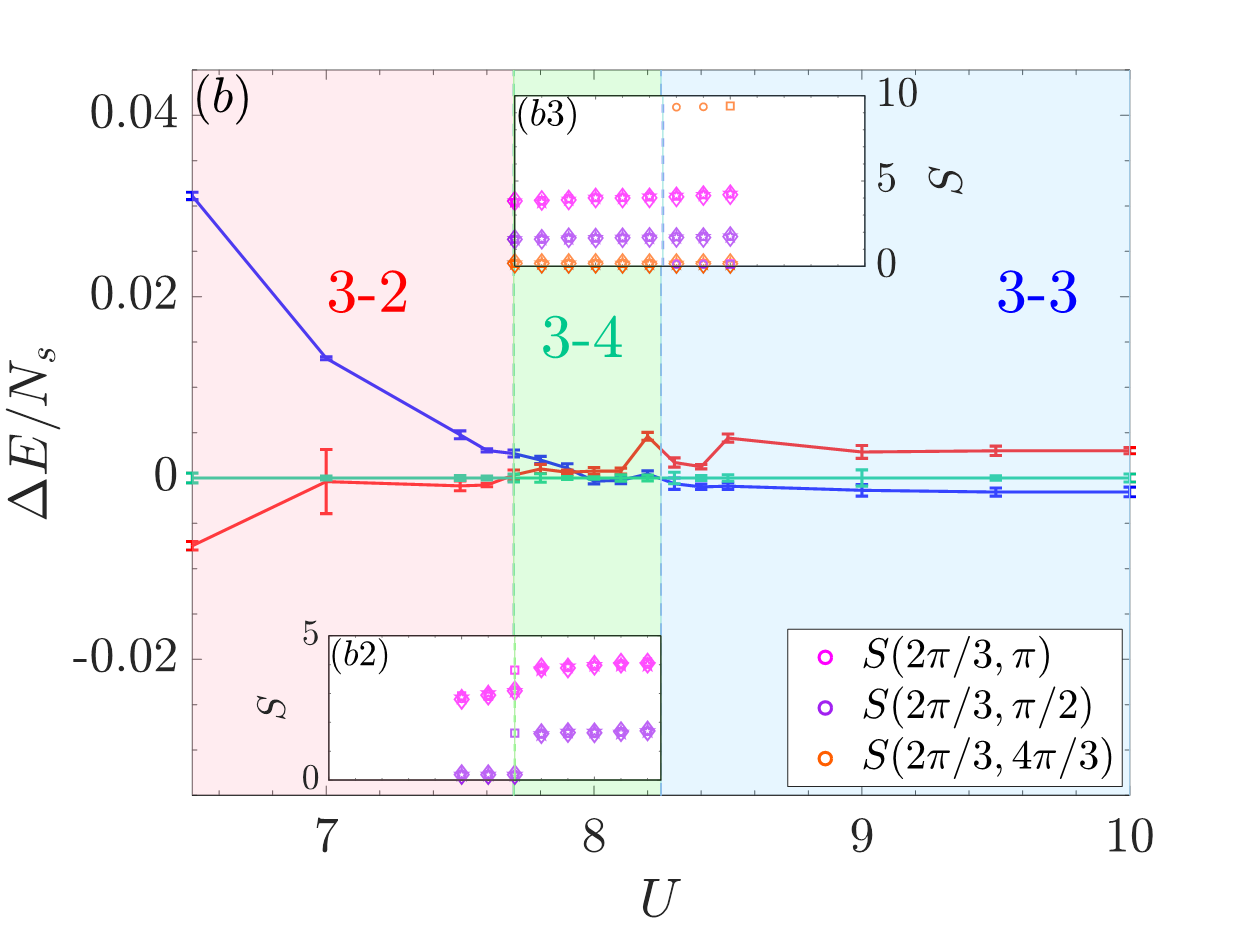}}
\caption{
Determining the transition points between phases. 
(a) The 3-2 AFM structure factor
vs.~$U$ 
is shown for three lattice sizes.
$S(2\pi/3, \pi)$ 
is small and nearly independent of the system size in the metallic phase
but grows proportionally to system size in the 3-2 AFM phase. The boundary between the two 
regimes gives 
the metal-insulator phase transition point $U_{c1}$.
The inset shows 
 $S(\pi,\pi)$ 
for $1/2$ filling, 
which exhibits a critical point at $U=0$, in contrast with the finite  $U_{c1}$ at $1/3$ filling.
 (b) Energy differences with respect to the 3-4 AFM phase. 
 Red, green and blue curves represent 
 3-2, 3-4, and 3-3 AFM states respectively. 
 Insets (b2) and (b3) show complementary results on structure factors to help 
 validate the location of $U_{c2}$ and $U_{c3}$. In each, the calculations use a trial wave function constructed from the superposition of the two states involved, and the 
 two corresponding structure factors are shown. 
Different symbol shapes represent results from 
different random seeds.
 } 
 \label{fig:combine-Sk-totalE}
 \end{figure*}

To further characterize the phase in different $U$ regions, the ground state structure factor at the three characteristic momenta 
$k=(2\pi/3, \pi/2), (2\pi/3, \pi), (2\pi/3, 4\pi/3)$ are shown in Fig.~\ref{fig:structure_factor} as functions of $U$.
The vertical lines 
separate the regimes where the different AFM orders are stable.
There is no peak in the PM metal;
in the 3-2 phase, a peak is present at ${k}=(2\pi/3, \pi)$;
in the 3-4 phase, peaks are seen at  both ${k}=(2\pi/3, \pi)$ and $(2\pi/3, \pi/2)$; 
finally the 
3-3 AFM phase shows a peak at ${k}=(2\pi/3, 4\pi/3)$. 
In a UHF calculation, the phase boundary can be determined by 
either searching for the global ground state at each $U$, or 
by comparing the energies of different self-consistent ordered solutions
--- ensuring in either cases 
that the thermodynamic limit is reached. 
This is challenging to execute in many-body computations where
statistical error bars and (much larger)
finite-size effects can exceed energy differences. We use a combination of 
strategies to resolve the different boundaries as illustrated in 
Fig.~\ref{fig:combine-Sk-totalE}.

The transition between the PM metal and the 3-2 AFM phases can be 
observed with robust and unambiguous self-consistent AFQMC calculations. 
In Fig.~\ref{fig:combine-Sk-totalE}(a), we show the computed
 structure factor $S(2\pi/3, \pi)$ as a function of $U$ 
 for a number of lattice sizes. At each $U$, we start the calculation 
 from a trial wave function generated from UHF with an essentially arbitrary $U_{\rm eff}$, and perform self-consistency via the natural 
 orbitals \cite{qin16}. The self-consistent iteration consistently converges 
 to the same final value (see Fig.~\ref{fig:self_consistency} in SM) which is shown in the plot. 
 The existence of the quantum critical point  $U_{c1}$ is evident from 
 the different size dependencies of $S(2\pi/3, \pi)$ as $U$ is varied across it. Extrapolating the finite- temperature compressibilities leads to a $U_{c1} $\cite{padilla2023} consistent with our results. 
A sharp contrast is seen between 
 this behavior and that at $1/2$-filling, shown in the inset of panel (a),
 emphasizing the result of decoupling of nesting from the Mott transition:
At $\langle n\rangle=3/2$,
the spins form an alternating pattern
 ``A(B\&C)A(B\&C)..." 
which singles out one species (e.g.~A) with the other two (B and C) behaving interchangeably \cite{honerkamp2004}.
 This spin arrangement has 
 period 2 in both the $x$ and $y$ directions, hence 
giving rise to a structure factor peak at $(\pi,\pi)$.  In this case, the dependence of the structure factor on lattice size starts  
 as soon as the interaction $U$ is turned on, implying that the long-range order develops at $ U_{\rm c1}=0$,
similar to the half filling case in the SU(2) Hubbard model.   This is expected since the SU(3) case also has
the same perfectly nested Fermi surface 
at ${k} = (\pi,\pi)$, as argued in \cite{honerkamp2004}.

To determine the nature of the ordered AFM phases, we apply 
complementary
measurements, with additional non-self-consistent calculations 
using multi-determinant trial wave functions. 
As discussed above, 
the UHF solutions using different $U_{\rm eff}$ can produce distinct 
patterns, which can be used as trial wave function and the initial Slater determinant to start our 
AFQMC calculations at any $U$. However, this turns the whole procedure from 
a linear process (projection) into a nonlinear process, which 
can get stuck in local minima. 
In this situation, we determine the true 
ground state by comparing the 
energies obtained from different constraints, as illustrated in 
panel (b) of Fig.~\ref{fig:combine-Sk-totalE}. 
Three regions are seen, with the lowest energies given by 
the 3-2, 3-4, and 3-3 states, respectively.
Their boundaries, where two energy curves cross, give two more transition points,
$ U_{c2}$ and $ U_{c3}$.
AFQMC with single Slater determinant constraints typically achieves energy accuracy of well over sub-percent 
level \cite{chang08,LeBlanc15,qin16_2,zheng17}. When the energy difference 
becomes much smaller, we augment our process ---
in the vicinity of $ U_{c2}$
we use a superposition of two Slater determinants corresponding to 3-2 \& 3-4 AFM   
as the trial wave function. 
This leads to 
a finite structure factor $S(2\pi/3, \pi)$ (3-2 AFM order) on one side, in contrast with
finite  $S(2\pi/3, \pi)$ and $S(2\pi/3, \pi/2)$ (3-4) on the other,
as illustrated in Fig.~\ref{fig:combine-Sk-totalE}(b2).
Near $U \sim U_{c3}$, we conduct the same procedures,
as shown in Fig.~\ref{fig:combine-Sk-totalE}(b3). 
Beyond $U/t\sim 8.25$, however, even this procedure leads to ambiguous results,
where both 3-4 
and 3-3 
structure factors can become finite, depending on the random number seed.
This combined with the tiny energy differences indicate that, to within the resolution of the current calculations, the 3-4 and 3-3 orders are essentially degenerate
for the larger $U$ values investigated here.


\vskip0.20in \noindent
\textit{Discussion.} 
Using high-accuracy many-body numerical computations, we have mapped out the 
ground-state phase diagram of the two-dimensional SU(3) Hubbard model at 1/3 filling (one particle per site). 
We find a metal-insulator phase transition 
at the quantum critical point $U_{c1} \sim 5.5$, 
above which long range magnetic orders develop.
Compared to the more ubiquitous  SU(2) counterpart,
the SU(3) case allows the separation of 
strong interaction effects at integer filling from non-generic features of band structure -- nesting of the Fermi surface and a van Hove singularity in the density of states.
An immediate consequence is the non-zero value of $U_c$. 

Above $U_{c1}$, our results suggest two novel intermediate magnetic orders (3-2 and 3-4 AFM phases)
before the final, strong coupling 3-3 phase.
We determined the critical interaction strengths for the ordering wave vectors by comparing their energies 
and also by using a superposition of two states as the initial and trial wave functions in our AFQMC calculations. The results given by these two approaches are consistent. Beyond $U_{c3} \sim 8.25$, the 3-3 AFM energy appears to be slightly lower than 3-4 AFM, but 
they are almost degenerate at the interaction strengths considered in the study.

The metal-insulator transitions and novel magnetic ordered phases predicted in our work provide important steps to understand SU($N$) physics.
This is especially timely with the intense experimental efforts to emulate such systems
using ultracold AEAs. 
How these magnetic orders evolve at finite temperature and other potentially exotic phases in the doped systems emerge are interesting questions for future work, especially in the experimental
context.

\vskip0.20in \noindent
\textit{Acknowledgements.} We thank Zewen Zhang for the discovery of the 3-4 order within Hartree-Fock calculations as well as useful discussions. E.V. acknowledges support from the National Science Foundation award number 2207048. 
Several calculations have been performed using the ACCESS-XSEDE allocation.  R.T.S. and E.I.G.P. are supported by the grant DE-SC-0022311, funded by the U.S. Department of Energy, Office of Science. K.R.A.H and E.I.G.P. acknowledge support from the Robert A. Welch Foundation (C-1872), the National Science Foundation (PHY-1848304), and the W. F. Keck  Foundation (Grant No. 995764). Computing resources were supported in part by the Big-Data Private-Cloud Research Cyberinfrastructure MRI-award funded by NSF under grant CNS-1338099 and by Rice University's Center for Research Computing (CRC). K.H.'s contribution benefited from discussions at the Aspen Center for Physics, supported by the National Science Foundation grant PHY-1066293, and the KITP, which was supported in part by the National Science Foundation under Grant No. NSF PHY-1748958. We thank the
Flatiron Institute Scientific Computing Center for computational resources. The Flatiron Institute is a division of the Simons Foundation.

\newpage
\centerline{\bf Supplemental Materials}
\vskip0.10in

 \vskip0.10in \noindent
In these supplemental materials, we present additional details concerning: (1) Self-consistency procedures we conduct in the constrained path Auxiliary Field Quantum Monte Carlo (CP-AFQMC) method; (2) Comparisons of energies in different phases and finite lattice size effect analysis in the Hartree-Fock.
\vskip0.10in \noindent
{\it Self consistency -- Real space spin correlations.}
In this paper we have
implemented a self-consistent CP-AFQMC approach, described in \cite{qin16}. 
An example of the convergence of the same-spin density-density correlations
  $\langle n_i^A n_j^A\rangle $ is given in Fig.~\ref{fig:self_consistency} to illustrate this process. 
The calculation is initiated from a Slater determinant given by the $U_{\rm eff}=3.5$ Hartree-Fock 3-2 AFM solution and the one-body density matrix is obtained from QMC. We diagonalize the one-body density matrix and find the eigenvectors which correspond to the
 largest $N_\sigma$ eigenvalues. Those eigenvectors are used to construct a new Slater determinant, which is then used as the 
initial Slater determinant for the next iteration of simulation. The number in the legend of Fig.~\ref{fig:self_consistency} 
indicates the self-consistent iteration value. Blue right triangles (``5" in the legend) and blue stars (``6") are almost on top of each other, implying the density-density correlations converge after $\sim 6$ self-consistent 
 iterations. 
 
 As a check that the results are independent of
 initialization, we also start our simulation with another initial Slater determinant, obtained from the $U_{\rm eff}=10,$ 3-2 AFM Hartree-Fock solution. The results 
 (red stars) agree with the one starting 
 from the $U_{\rm eff}=3.5$ Hartree-Fock (blue star).  
 All results presented in the main text are given 
 by the last self-consistent (fully converged) iteration.

\begin{figure}[h]
\includegraphics[width=1\columnwidth]{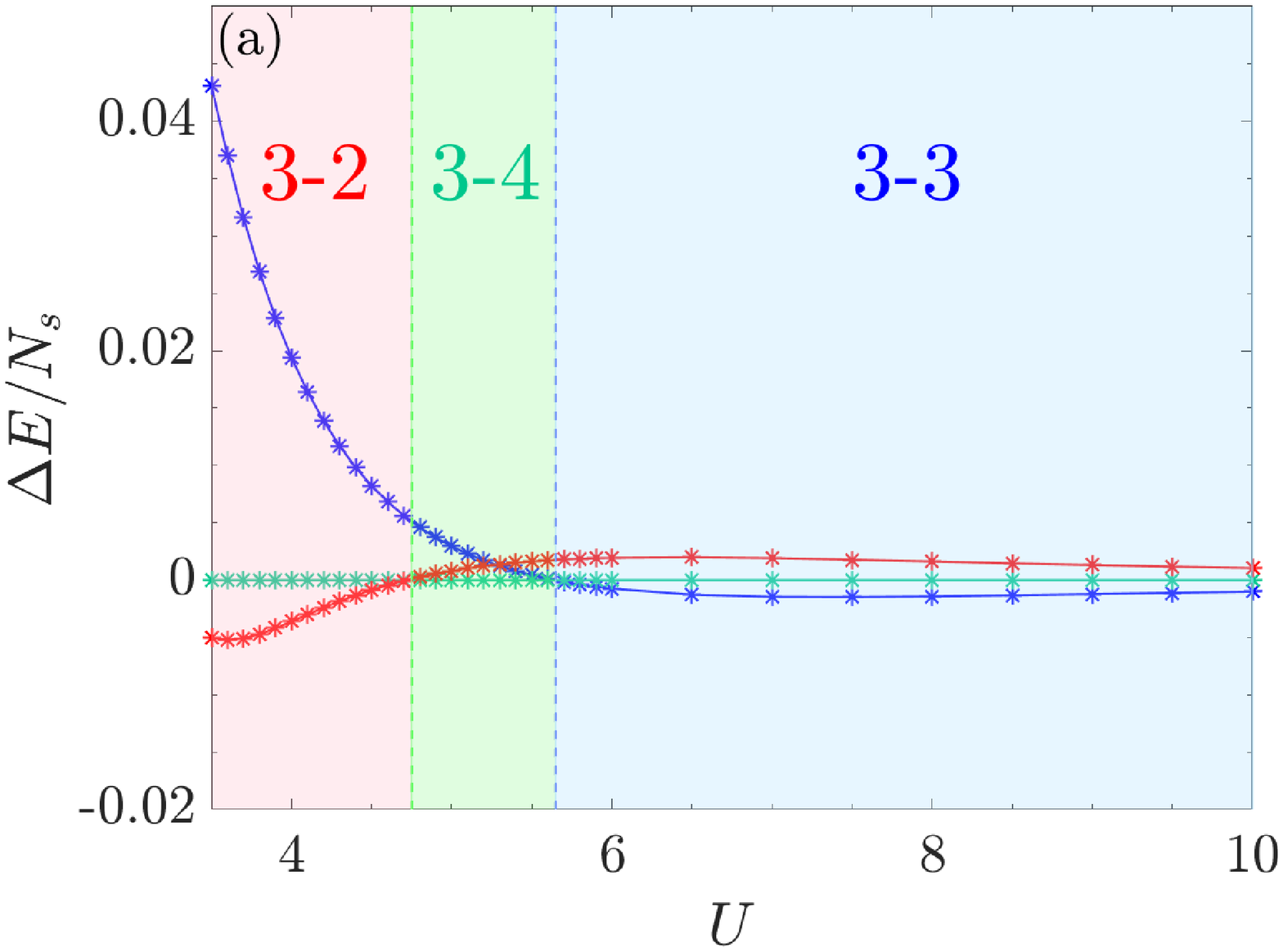}
 \includegraphics[width=1\columnwidth]{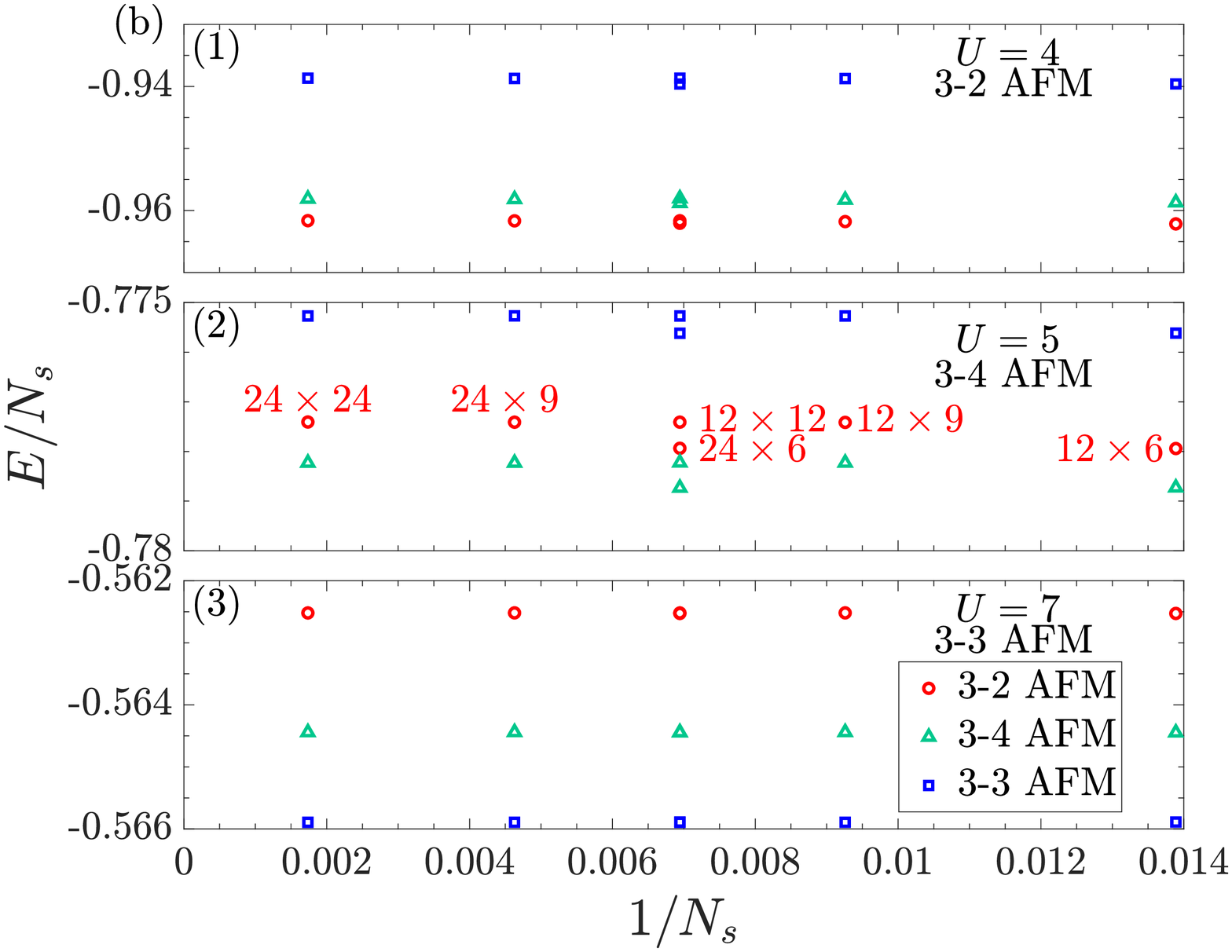}
 
\caption{(a) The energy difference $\Delta E=E-E_{\rm 3-4 AFM}$ (3-4 AFM energy is viewed as a reference) per site given by Hartree-Fock as a function of interaction strength. 3-2, 3-4, 3-3 AFM energies are depicted by red, green and blue curves respectively. (b) Energy per site $E/N_s$ vs. $1/N_s$, given by the Hartree-Fock calculation for $U=4, 5, 7$, corresponding to ``3-2", ``3-4" and ``3-3" AFM phases respectively. The lattice sizes $L
_x \times L_y$ are labeled next to the data points in panel (2) as examples. Different symbols in the legend represent the possible ``local minimum" orders.}
 \label{fig:SM_MFT_lattice_size_effect}
 \end{figure}
 
\vskip0.10in \noindent
{\it Finite lattice size effects --   Energy.}
In Fig.~\ref{fig:SM_MFT_lattice_size_effect}, we explore finite lattice size effects in the Hartree-Fock calculations. First, similar to Fig.~\ref{fig:combine-Sk-totalE}(b), we show the total energy difference per site $\Delta E/N_s$ given by Hartree-Fock as a function of $U$ for a $12 \times 12$ square lattice in panel (a). The 3-2, 3-4, and 3-3 AFM phase has the lowest energy for $U_{c1}^{MF} \lesssim U \lesssim U_{c2}^{MF}$, $U_{c2}^{MF} \lesssim U \lesssim U_{c3}^{MF}$ and $U \gtrsim U_{c3}^{MF}$ respectively. In order to detect whether the small energy difference at strong interactions is due to the finite lattice size effect, we choose $U=4$ (3-2 phase), $U=5$ (3-4 phase) and $U=7$ (3-3 phase)
and present the energy per site $E/N_s$ as a function of $1/N_s$ in Fig.~\ref{fig:SM_MFT_lattice_size_effect}(b). Different symbols in the legend represent the local minimum 3-2, 3-4, 3-3 AFM states. Lattice sizes are labeled next to the data points in Fig.~\ref{fig:SM_MFT_lattice_size_effect}(b)-(2)). We can extrapolate the energy to the thermodynamic limit $1/N_s \to 0$. There is no crossing between different AFM phases, implying the energy ordering between the three magnetic phases on the $12 \times 12$ lattice in Fig.~\ref{fig:SM_MFT_lattice_size_effect}(a) still holds in the thermodynamic limit.  We have also verified that this conclusion is robust to the application twisted boundary conditions 
to the lattice and taking the average of energies for different twist angles.  This method effectively samples a finer mesh of momentum points and hence larger spatial sizes\cite{sorella2015finite,gros1996control,lin2001twist}.

\bibliography{CPQMC_SU3_bib}

\end{document}